# Uranium Doping and Thermal Neutron Irradiation Flux Pinning Effects in MgB$_2$

T. Silver[1*], J. Horvat[1], M. Reinhard[2], Y. Pei[3], S. Keshavarzi[1], P. Munroe[4], and S.X. Dou[1]

[1]*Institute for Superconducting and Electronic Materials, University of Wollongong, Northfields Avenue, Wollongong 2522, Australia*
[2] *Safety and Radiation Science, Australian Nuclear Science and Technology Organisation (ANSTO), Lucas Heights 2234, Australia*
[3]*Centre for Electron Microscope Analysis, TienJun University, TienJun, PRC*
[4]*Electron Microscopy Unit, University of New South Wales, Sydney 2052, Australia*

## ABSTRACT

The U/n method is a well-established means of improving flux pinning and critical current performance in cuprate superconductors. The method involves the doping of the superconductor with $^{235}$U followed by irradiation with thermal neutrons to promote fission. The resultant columnar damage tracks produced by the energetic fission products pin flux vortices and improve critical current performance in magnetic fields. No such improvement could be observed when the U/n method was applied to MgB$_2$ superconductor. No fission tracks could be observed in TEM, even for samples that were irradiated at the highest fluence. Gamma-ray spectroscopy indicated that fission had occurred in the expected way.

## 1. INTRODUCTION

Since its superconducting properties were first discovered in 2001 [1, 2], MgB$_2$ has been found to share characteristics with both conventional low temperature superconductors (LTS) and with cuprate high temperature superconductors (HTS). Its critical temperature T$_c$ is nearly 40K, unusually high for a non-cuprate material. MgB$_2$ behaves like a conventional type-II superconductor in relation to the isotope effect, the T-dependence of the upper critical field, and resistivity R(T) measurements, but more like the HTS superconductors in the temperature dependence λ(T) of the penetration depth and the behaviour of the Hall coefficient near T$_c$ [3]. Although MgB$_2$ films have attained critical currents above $10^6$ A/cm$^2$ at 4.2K in low magnetic fields [4], MgB$_2$ [3] and the HTS cuprates [5] share a rapidly decreasing J$_c$(H) performance as magnetic fields increase compared to LTS superconductors such as Nb$_3$Sn and Nb-Ti. In HTS materials the field performance is related to both weak links between grains and poor flux pinning behaviour. The former consideration does not apply to MgB$_2$, as transport measurements of J$_c$ and measurements calculated from magnetic hysteresis loops yield very similar results, indicating that the flow of super-current is not hindered by grain boundaries [6, 7]. The flux creep in MgB$_2$ is much weaker than in average HTS [8, 9], indicating stronger vortex pinning. However, this pinning should still be improved to offset the effect of the low H$_{c2}$ of MgB$_2$ (about 18T) [10] on the field dependence of J$_c$.

---

* Corresponding author: Tania M. Silver, Institute for Superconducting and Electronic Materials, Faculty of Engineering, University of Wollongong, Northfields Ave., Wollongong, NSW 2522, Australia; telephone: (02) 4221 5766; fax: (02) 4221 5731; e-mail: tsilver@uow.edu.au.

Superconductivity above $H_{c1}$ in any type-II material depends on flux pinning in regions where the superconducting order parameter is reduced, either by intrinsic features of the crystal structure or by point or extended defects. Otherwise the flux vortices would move individually or collectively due to Lorentz forces when a uniform current is applied, resulting in local superconducting phase slip and a non-zero electrical resistance [11]. There is a long history of attempts, often successful, to improve $J_c(H)$ performance by creating strong pinning centres in HTS superconductors, by means such as the introduction of precipitates and the use of a variety of irradiation techniques employing neutrons, heavy ions, electrons and protons [5]. Some methods of introducing pinning centres have also been successful with $MgB_2$, including oxygen alloying in thin films [4] and the addition of nano-scale particles of SiC [12] and Si [13], as well as irradiation with protons [14]. Attempts have also been made using irradiation with heavy ions [15, 16] and neutrons [17, 18], although reported effects from these two methods have been small to date.

One irradiation method that has not yet been reported for $MgB_2$ is the U/n [19, 20] method, although it has been highly successful in improving flux pinning in HTS. The U/n method differs from straightforward irradiation with thermal neutrons, in that the material is first doped with compounds containing $^{235}U$. When irradiation takes place the $^{235}U$ atoms absorb thermal neutrons and fission. The two fission products recoil in opposite directions with a total kinetic energy of approximately 160 MeV, creating extended defects in HTS in the form of two fission tracks. Fission tracks are made up of amorphous material and are approximately 10 µm long and 10 nm in diameter in Bi-based HTS superconductors [21]. The track structure is discontinuous and randomly oriented. U/n achieves its greatest success with highly anisotropic Bi-based (BSCCO) superconductors such as Bi-2223/Ag tape, where a 500-fold improvement in $J_c$ was reported [22] at 77K and 0.8T for H//c. $J_c$ for H//ab under the same conditions was nearly four orders of magnitude larger than for H//c before irradiation. There was also improvement after irradiation for H//ab, but $J_c$ improved less than one order of magnitude. The anisotropy $J_c(H//ab)/J_c(H//c)$ was reduced 23 times at 0.5T and 77K. A 20-fold improvement in $J_c$ at 77K and 0.25 T has been reported [23] in bulk melt textured $YBa_2Cu_3O_{7-\delta}$ (YBCO).

The discrepancy between the results for the YBCO and the Bi-2223 tape is due to the lesser anisotropy of YBCO. The vortices in HTS are two dimensional (2D) pancake vortices, residing on the $CuO_2$ planes [24, 25]. In highly anisotropic HTS, these pancake vortices in each $CuO_2$ layer can move independently from the vortices in the other $CuO_2$ layers. This makes it difficult to introduce effective pinning centres into such types of HTS, because vortices in each of the layers would have to be pinned down, giving an unacceptably high density of defects. However, if the pinning centres are in the shape of long columns traversing the $CuO_2$ planes, the pancake vortices in different layers will be collectively pinned along these columns and will thus be effectively aligned into vortex lines. Such pinning has been shown to be much more effective than pinning by point-like pinning centres [5].

The U/n method would seem to also be appropriate for $MgB_2$. The reported coherence length values [3] of $\xi_{ab}(0) = 3.7-12$ nm and $\xi_c(0) = 1.6-3.6$ nm are of the same order as the expected diameter of the fission tracks introduced by high energy fragments resulting from the fission of $^{235}U$. While it has been generally accepted that the optimal size of a pinning centre is twice the coherence length, there is also a theoretical indication that larger sizes up to the penetration depth $\lambda$ may be effective under some circumstances [26]. Pinning centres of the size of $\lambda$ would be effective if the pinning were dominated by electromagnetic interaction of vortices with pinning centres [27,28].

The U/n method would not be expected to produce anything like a 500-fold improvement in $J_c$ in $MgB_2$ because the material is far less anisotropic than BSCCO. Reported results for the anisotropy coefficient $\gamma = \xi_{ab}/\xi_c$ range from 1.1 to 2.7 for textured bulk, aligned crystallites, films and single crystals [3, 29]. By comparison, γ has been reported as 30 for Bi-2223/Ag tape [30] and 3 for melt-textured bulk YBCO [31]. There is no evidence of a 2D vortex state in $MgB_2$ bulk material, and the I-V characteristics have been reported as consistent with a vortex glass model [32, 33].

The most serious problem for the U/n method in $MgB_2$ is the enormous cross section of 3837 barn (b, where $1b = 10^{-28}$ $m^{-2}$) for the $^{10}B(n,\alpha)^7Li$ nuclear reaction. Here a $^{10}B$ nucleus and a thermal (of the order of 25 meV) neutron react to form an α particle of kinetic energy 1.47 MeV and a $^7Li$ nucleus with a kinetic energy of 0.84 MeV. By comparison, the $^{235}U$ fission cross section is 584 b. Natural boron is comprised of 19.9% $^{10}B$, with the balance $^{11}B$. The $^{10}B(n,\alpha)^7Li$ reaction was deliberately used by Babic et al. [17] in an attempt to improve flux pinning by introducing ion tracks into $MgB_2$. An enhancement in the upper critical field was observed. However, the distribution of defects would have been extremely inhomogeneous because the thermal neutrons in this material have a mean free path of ~ 0.2 mm and thus only penetrated into a thin surface layer before being completely absorbed. The penetration depth increases with higher neutron energy, as the cross section for the $^{10}B(n,\alpha)^7Li$ reaction decreases. At a neutron energy of 14 MeV the reaction cross section is 48.95 mb. Fast neutrons can create defects directly via elastic and inelastic scattering with the atomic lattice. However, defects produced by these mechanisms are considerably less significant than those produced via the $^{10}B(n,\alpha)^7Li$ reaction when a full spectrum of neutron energies is employed [18]. In using the U/n method, it should be possible to significantly eliminate the competition for thermal neutrons between $^{10}B$ and $^{235}U$ by using highly enriched $^{11}B$ instead of natural boron. $^{11}B$ has a total thermal neutron cross-section of 5.05 b, almost all of it representing elastic scattering.

## II. EXPERIMENTAL

To test the applicability of the U/n technique for enhancing $J_c$ in $MgB_2$ an experimental program was designed. Powders of 1-11μm Mg from Hypertech (for samples A-D) or 325 mesh Mg (for sample E) and crystalline B (99.5 at% $^{11}B$, particle size <22 μm, from Eagle-Picher) were mixed in a mortar in a stoichiometric atomic ratio of 1:2. Half of the powders were mixed with $UO_2$ (93% enriched in $^{235}U$) to give 1 wt% U. Optical microscopy revealed that the $UO_2$ particles ranged from 2-4 μm in diameter.

The powders were pressed into 0.4 g pellets, and the pellets sealed in iron tubes. Sintering took place in a tube furnace under flowing argon. For samples A-D the temperature was ramped up to 760° C over one hour. The pellets were held at that temperature for 30 min and then furnace cooled. For sample E the temperature was ramped up to 900° C over 1.5 h, and the pellets were held at that temperature for 3 h, then furnace cooled. The densities of the samples were 1.4 g/cm$^3$ (samples A-D) and 1.0-1.1 g/cm$^3$ (sample E). X-ray diffraction (XRD) was used to determine the phase composition of the doped and un-doped samples, while Energy Dispersive Spectroscopy (EDS) was used to determine the uranium distribution of doped samples. Portions of the doped and un-doped samples were then cut and filed into small regular rectangular blocks with dimensions of typically 3-4 × 3 × 2 mm$^3$. The magnetisation and ac susceptibility were determined using a Quantum Design Physical Properties Measurement System (PPMS) in a time varying magnetic field with sweep rate 50 Oe/s and amplitude up to 8.5 T, with the applied magnetic field parallel to the longest dimension of the sample. The low field Zero Field Cooled (ZFC) and Field Cooled (FC) measurements were made with a Quantum Design Magnetic Properties Measurement System (MPMS). The $J_c$ values at different magnetic fields and temperatures were calculated from the magnetic hysteresis loops using the Bean model: $J_c=20\Delta M/[a(1-a/3b)]$, where *a* and *b* are the dimension of the sample perpendicular to the applied field, *a<b*. $T_c$ was determined from the ac susceptibility measurements.

Small block samples, in doped and un-doped pairs, were double encapsulated in titanium capsules and irradiated by neutrons in HIFAR, a 10 MW DIDO class research reactor operated by ANSTO at Lucas Heights, Australia. The irradiation rig was located within the graphite region of the reactor, and thus the neutron energy spectrum was highly thermalised with only a minimal fast and epithermal neutron component. The rig was rated with a nominal neutron flux of ~$1\times10^{13}$ n·cm$^{-2}$s$^{-1}$. Four different fluences were used: $2 \times 10^{16}$cm$^{-2}$ (A), $5 \times 10^{15}$cm$^{-2}$ (B, E), $5 \times 10^{14}$cm$^{-2}$ (C), and $5 \times 10^{13}$cm$^{-2}$(D). Following irradiation, the J$_c$ and flux pinning properties of the irradiated pellets were then compared with the pre-irradiated results. The doped, irradiated samples were examined by transmission electron microscopy (TEM) to detect any fission tracks, and the extent of fission determined by gamma ray spectroscopy.

### III.      RESULTS AND DISCUSSION

X-ray diffraction (XRD) using a Philips 1730 phase XRD diffractometer showed that greater phase purity could be obtained using the longer, hotter sintering to fully react the crystalline boron and the magnesium (see Fig. 1). After 3 h at 900 C, the sintered un-doped sample E material (a) is MgB$_2$ with traces of Mg and MgO. The sintered 1wt% U-doped sample E material (b) also consists of MgB$_2$ with traces of Mg, MgO and UO$_2$. Fig. 1(c) shows the un-doped sintered material characteristic of samples A-D. Because the sintering was shorter and at lower temperature the Mg and MgO impurity contents are much higher, due to the much larger amount of remaining un-reacted crystalline boron. These short sintering conditions are appropriate when amorphous boron is used to make MgB$_2$. The impurities would be expected to give a smaller superconducting volume, but the intra-granular properties affected by the U/n method would not be degraded. For this reason samples from this lower quality material could also be used to determine the optimum neutron fluence.

Samples were characterised using a JEOL JXA-840 scanning electron microscope (SEM) equipped with a Link Systems AN10000 energy dispersive spectrometers (EDS). The results showed MgB$_2$ grains in a porous structure. The UO$_2$ particles remained separate, and the larger agglomerates could be distinguished at the grain boundaries. EDS analysis revealed the presence of large amounts of Mg and small amounts of U and Fe, the source of which was most likely the iron tubing containing the samples during sintering. The boron could not be detected with the experimental set-up. Figure 2(a) shows an EDS spectrum of the doped version of sample E. EDS mapping revealed a reasonably even distribution of UO$_2$ particles among the MgB$_2$ grains. Figure 2(b) shows the fairly even distribution of small UO$_2$ particles throughout the MgB$_2$ in sample E.

The ac susceptibility of the samples was measured at temperatures from 5 to 45 K and the magnetisation from 5 to 30 K using a Physical Properties Measurement System (Quantum Design, San Diego). T$_c$ was determined from the ac susceptibility, and was not found to be significantly affected by irradiation and doping. T$_c$ was measured to be 38.5 K with a sharp transition for the higher quality sample E regardless of doping and approximately 37.2 K for the other samples. The discrepancy is probably because samples A-D have a greater concentration of non-magnetic impurities, which are known to depress T$_c$ in MgB$_2$ [34]. Figure 3(a) shows the derived values of the critical current at 5, 20 and 30K as a function of magnetic field for three versions of sample A ((i) un-doped and un-irradiated, (ii) doped with 1 wt% U, but not irradiated, and (iii) the same piece of doped sample A after irradiation with thermal neutrons at $2 \times 10^{16}$ cm$^{-2}$) and for sample B (doped with 1% U, then irradiated with thermal neutrons at $5 \times 10^{15}$ cm$^{-2}$). J$_c$ values for sample E were similar, with a slightly lower H$_{irr}$, probably because the lower density cancelled out any benefits due to the higher purity. It is impossible to accurately determine J$_c$ at 5K for very low fields because of the presence of thermal flux jumping [35]. J$_c$ performance is slightly worse in the doped samples except at high magnetic fields, probably due to the presence of the additional impurities, but there are no significant differences between the irradiated and un-irradiated samples, in contrast to what has been demonstrated for HTS superconductors. The slightly worse performance of the high fluence irradiated sample A at high fields may be due to additional radiation damage from the higher neutron fluence.        The discrepancy between the magnetisation for ZFC and FC magnetisation measurements under superconducting conditions provides a measure of the pinning. Fig. 3(b) shows the ZFC-FC results at 100 Oe as a function of temperature for the doped, irradiated versions of samples A-D. There are no significant differences with fluence. Fig. 3(c) shows the pre-irradiation ZFC-FC results for the doped and un-doped versions of sample E. The higher $T_c$ is attributed to the greater phase purity.

Transmission electron microscopy (TEM) was performed on two specimens (A and B) doped to 1 wt% U and irradiated to neutron fluences of $2 \times 10^{16}$ cm$^{-2}$ and $5 \times 10^{15}$ cm$^{-2}$ respectively. For both materials the microstructure was similar, that is, there were grains of $MgB_2$ several microns in diameter in which a number of dislocations and other crystalline defects were evident, but no fission tracks. Many such grains were examined, and Fig. 4(a) shows a typical example. This observation is in contrast to TEM studies of U-doped Bi-based HTS specimens exposed to similar levels of irradiation, in which the fission tracks appear as straight, randomly-oriented black lines due to the greater electron scattering from the amorphous material they contain. These fission tracks are several microns in length and clearly visible in TEM even at modest magnifications. This is illustrated in Figure 4(b), which shows a TEM image of the core of an irradiated, uranium-doped Bi-2223/Ag tape. A quantitative analysis of $^{235}U$ fissions in the present $MgB_2$ samples was obtained from the fission product yield as measured by the gamma-ray spectrometric method [36]. Measurements were limited to the doped, irradiated versions of samples C and D. A period of 92 days had expired since irradiation, which meant that most of the short-lived radioisotopes had decayed. This simplified the spectrum. To facilitate measurements, samples were placed 10 cm from the face of a large volume coaxial high purity germanium (HPGe) detector.

The gamma ray spectrum for sample C is shown in Figure 5. The spectral features were consistent with a 96 day decayed $^{235}U$ fission product spectrum [37]. The actual fission yield (see Table 1) was determined from the measured $^{137}Cs$ activity and the cumulative fission yield data for $^{137}Cs$ from $^{235}U$. The cumulative yield includes the total number of $^{137}Cs$ atoms produced directly as a result of fission in addition to $^{137}Cs$ atoms produced via radioactive decay of the $^{137}Cs$ precursors, namely $^{137}I$ and $^{137}Xe$ which both have half-lives of less than 5 minutes. By comparison the half-life of $^{137}Cs$ is 30.17 years. The $^{137}Cs$ activity was determined from the peak area of the 662 keV emission and the known detector efficiency at this energy.

The fission yield can also be estimated from knowledge of the sample $^{235}U$ content, the cross section for fission of $^{235}U$ and the neutron fluence. The estimated fission yield along with the results from the gamma-ray spectrometric method is given in Table 1. A good correlation between estimated and measured fission yield was obtained. This shows that nuclear fission of $^{235}U$ did occur as expected. In HTS the same fission would produce columnar defects (Fig. 4(b)), resulting in strong pinning improvement. However, not a single fission track was observed for $MgB_2$ (Fig. 4(a)), and the field dependence of $J_c$ was virtually unchanged after the irradiation (Fig. 3(a)).

An important parameter defining the creation of fission tracks is the electronic stopping power, also called the electronic energy loss, $S_e$ [38]. It is defined as the energy transfer into the electronic excitations of the target atoms per unit length of the ion path through the target crystal. In our experiments, heavy ions of about 80 MeV are created by the nuclear fission and transfer their energy to the $MgB_2$. The fission tracks can occur only if $S_e$ exceeds a threshold value $S_{e0}$ [39], which is dependent on the target material and the properties of the products of fission. The occurrence or otherwise of the fission tracks can be described by the thermal spike model [40, 41, 42]. In order to create fission tracks, there has to be a mechanism of energy transfer from the electrons to the crystal lattice. The thermal spike model does not explain the nature of these interactions, however, it explains the formation of the columnar defects in a thermodynamic approach. Because of the transfer of energy from the heavy ions to the electrons and subsequently to the crystal lattice on a longer time scale, a sudden localised increase in temperature occurs along the ion path through the crystal. For values of $S_e$ higher than the threshold value $S_{e0}$, the crystal lattice melts in a very localised volume along the ion track. Because the diameter of the molten volume is very small (of the order of a nanometre), this heat is diffused through the crystal lattice quickly enough to freeze the molten volume without allowing re-crystallisation. Consequently, a track of amorphous material is formed in the crystal, which is not superconducting and is expected to be a strong pinning centre.

In our experiments the fission tracks were clearly not formed. The reason for this may be that the $S_e$ values were too low for the array of fission products interacting with $MgB_2$, or that the heat conductivity required for freezing the molten volume is too low in $MgB_2$, or that the energy transfer from the ions to the crystal electrons and the crystal lattice does not occur on timescales that would enable melting of the lattice. Because all of these factors are still unknown for $MgB_2$, we cannot ascribe the observed lack of fission tracks to any of them in particular.

It is interesting to compare our results with experiments employing 2 MeV proton irradiation [14] as a means of improving the vortex pinning in $MgB_2$. There, an improvement of the field dependence of $J_c$ and the irreversibility field was observed. In HTS, protons of these energies were not capable of producing amorphous columnar defects, which are similar in nature to fission tracks [43]. To achieve such columnar defects, 800 MeV protons had to be used [44]. Therefore, the 2 MeV protons used in Ref. 14 were producing point defects in $MgB_2$, which are much weaker pinning centres than columnar defects. However, due to the very large density of defects introduced, an observable increase of pinning was obtained. Point defects are also likely to be responsible for the improvement in pinning reported in Ref. 17 due to the $^{10}B(n,\alpha)^7Li$ nuclear reaction, rather than any ion tracks. Our results show a signature of point defects as well. There is a slight improvement of the field dependence of $J_c$ at high fields for the highest neutron fluences (see Fig. 3(a)), accompanied by a slight decrease of $J_{c0}$ and $T_c$. This may have occurred due to the point defects introduced by neutrons, as well as by the fission products. Because the neutron fluence employed here was much smaller than the proton fluence in Ref. 14, the improvement in the pinning was correspondingly smaller.

Confirmation of the nature of the crystallographic damage produced by both 2 MeV protons and $^{235}U$ fission particles in $MgB_2$ was obtained using particle transport simulation software [45]. Fission particles were simulated using $^{137}Cs$ ions with a kinetic energy of 60 MeV. Results revealed that the typical range of a 2 MeV proton in $MgB_2$ is about 45 μm. This can be compared to the typical range of a $^{137}Cs$ ion in $MgB_2$ of 10.6 μm. In terms of the number of Mg or B recoils produced, 2 MeV protons produced on average 25, as opposed to ~33,000 in the case of the $^{137}Cs$ ions. In terms of the damage structure, protons produced small damage regions consisting of isolated point defects localised about the end of the proton range. In comparison, the $^{137}Cs$ ions produced large damage clusters containing several hundred displaced atoms. The clusters were located along the track of the ions, although most damage was again located within the last 2-3 μm of the track. In both cases annealing of the displacement damage immediately following creation could be expected due to thermally assisted diffusion of atoms back to their correct crystallographic sites. The thermal energy assisting this process is sourced directly from the interacting primary and secondary particles produced in the actual displacement event. Longer term annealing may also be expected due to room temperature assisted processes. The fraction of annealing occurring within the fission particle damaged material is expected to be greater than that within the proton irradiated material. This is on account of the increased density of damage within the clusters of the fission particle damaged material as opposed to the more isolated point defects predicted in the proton-irradiated material. The damage remaining in the fission particle damaged material is still expected to be much greater than that found in the proton-irradiated material.

## IV. SUMMARY

MgB$_2$ samples were made of highly enriched $^{11}$B in a reaction in-situ process and doped with UO$_2$ containing highly enriched $^{235}$U to a doping level of 1 wt% U. The samples were then irradiated with thermal neutrons at fluences ranging from $5 \times 10^{13}$ cm$^{-2}$ to $2 \times 10^{16}$ cm$^{-2}$. The intent was that the $^{235}$U atoms would fission and that the fission fragments would create columnar amorphous damage tracks to act as pinning centres to trap magnetic flux and improve critical current performance in the MgB$_2$ superconductor. There was no evidence of such tracks in TEM. This is consistent with the lack of evidence of improved critical currents in magnetic fields as a result of the irradiation. Gamma-ray spectroscopy indicated that fission of the $^{235}$U atoms had taken place as expected and was not hindered by the presence of any residual $^{10}$B. The reason for the absence of the tracks is not yet clear, because the threshold electron stopping power and the mechanisms of the energy transfer to the crystal lattice are still not known for the interaction of $^{235}$U fission fragments with MgB$_2$.

## ACKNOWLEDGEMENTS

We wish to thank X.L. Wang, K. Konstantinov, A. Pan, M.J. Qin, M. Ionescu, and B.R. Winton of the Institute for Superconducting and Electronic Materials and D. Wexler of the Faculty of Engineering, University of Wollongong for practical help and useful discussions. The authors would also like to thank the Australian Institute of Nuclear Science and Engineering for providing assistance (Award No. 02/121) to enable the work on MgB$_2$ to be conducted.

[1] J Akimitsu 2002 *Symposium on Transition Metal Oxides (Sendai, Japan, 10 January 2001)*.

[2] J. Nagamatsu, J. Nakagawa, T. Muranaka, Y. Zenitani and J. Akimitsu, *Nature* **410** (2001), 63.

[3] C. Buzea and T. Yamashita, *Supercond. Sci. Technol.* **14** (2001) R115.

[4] C.B. Eom, M.K. Lee, J.H. Choi, L.J. Belenky, X. Song, L.D. Cooley, M.T. Naus, S. Patnaik, J. Jiang, M. Rikel, A. Polyanskii, A. Gurevich, X.Y. Cai, S.D. Bu, S.E. Babcock, E.E. Hellstrom, D.C. Larbalestier, N. Rogado, K.A. Regan, M.A. Hayward, T. He, J.S. Slusky, K. Inamaru, M.K. Haas and R.J. Cava, *Nature,* 2001, **411**, 558.

[5] M.E. McHenry and R.A. Sutton, *Progress in Materials Science* **38** (1994) 159.

[6] K.H.P. Kim, W.N. Kang, M.S. Kim, C.U. Jung, H.J. Kim, E.M. Choi, M.S. Park and S.I. Lee, *Physica C* **370** (2002) 13.

[7] Y. Bugoslavsky, G.K. Perkins, X. Qi, L.F. Cohen and A.D. Caplin, *Nature* **410** (2001) 563.

[8] H.H. Wen, S.L. Li, Z.W. Zhao, H. Jin, Y.M. Ni, W.N. Kang, H.J. Kim, E.M. Choi and S.I. Lee, *Phys. Rev. B* **64** (2001) 134505.

[9] H.H. Wen, S.L. Li, Z.W. Zhao, H. Jin, Y.M. Ni, Z.A. Ren, G.C. Che and Z.X. Zhao, *Physica C* **363** (2001) 170.

[10] D.C. Larbalestier, L.D. Cooley, M.O. Rikel, A.A. Polyanskii, J. Jiang, S. Patnaik, X.Y. Cai, D.M. Feldmann, A. Gurevich, A.A. Squitieri, M.T. Naus, C.B. Eom, E.E. Hellstrom, R.J. Cava, K.A. Regan, N. Rogado, M.A. Hayward, T. He, J.S. Slusky, P. Khalifah, K. Inamaru and M. Haas, *Nature* **410** (2001) 186.

[11] D.S. Fisher, M.P.A. Fisher and D.A. Huse, *Phys.Rev.B* **43** (1991) 130.

[12] S.X. Dou, S. Soltanian, J. Horvat, X.L. Wang, S.H. Zhou, M. Ionescu and H.K. Liu, *Appl.Phys.Lett.* **81** (2002) 3419.

[13] X.L. Wang, S.H. Zhou, M.J. Qin, P.R. Munroe, S. Soltanian, H.K. Liu and S.X. Dou, cond-mat/0208349, to be published in *Physica C*.

[14] Y. Boguslavsky, L.F. Cohen, G.K. Perkins, M. Polichetti, T.J. Tate, R. Gwillam and A.D. Caplin, *Nature* **411** (2001) 561.

[15] H. Narayan S.B. Samanta A. Gupta, A. V. Narlikar, R. Kishore K.N. Sood, D. Kanjilal, T. Muranaka and J. Akimitsu, *Physica C*. **377** (2002) 1.

[16] R.J. Olsson, W.-K. Kwok, G. Karapetrov, M. Iavarone, H. Claus, C. Peterson and G.W. Crabtree, cond-mat/0201022.

[17] E. Babic, B. Miljanic, K. Zadro, I. Kusevic, Z. Marohnic, D. Drobac, X.L. Wang and S.X. Dou, *Fizika A* **10** (2001) 87.

[18] M. Eisterer, M. Zehetmayer, S. Tonies and H.W. Weber, *Supercond. Sci. Technol.* **15** (2002) L9.

[19] R. Weinstein, Y. Ren, J. Liu, I. Chen, R. Sawh, C. Foster and C. Obot, *Proc. Int. Symp. on Superconductivity, Hiroshima, 1993.* (Springer, Berlin, 1993) p. 285.

[20] G.W. Schulz, C. Klein, H.W. Weber, S. Moss, R. Zeng, S.X. Dou, R. Sawh, Y. Ren and R. Weinstein, *Appl.Phys.Lett.* **73** (1998) 3935.

[21] S. Tönies, H.W. Weber, Y.C. Guo, S.X. Dou, R. Sawh and R. Weinstein, *Appl. Phys. Lett.* **78** (2001) 3851.

[22] S.X. Dou, Y.C. Guo, D. Marinaro, J.W. Boldeman, J. Horvat, P. Yao, R. Weinstein, A. Gandini, R. Sawh and Y. Ren, *IEEE Trans.Appl.Supercond.* **11** (2001) 3896.

[23] M. Eisterer, S. Tonies, H.W. Weber, R. Weinstein, R. Sawh and Y. Ren, *Physica C* **341-348** (2000) 1439.

[24] J. R. Clem, *Phys. Rev. B.* **43** (1991) 7837.

[25] J.R. Clem, *Physica C* **162-164** (1989) 1137.

[26] N. Takezawa and K. Fukushima, *Physica C* **290** (1997) 31.

[27] G.S. Mktrchyan and V.V. Shmidt, *Sov. Phys. JETP* **34** (1972) 195.

[28] W.E. Timms and D.G. Walmsley, *Phys. Stat. Solidi B* **71** (1975) 741.

[29] A. Dulcic, M. Pozek, D. Paar, E.M. Choi, H.J. Kim, W.N. Kang and S.I. Lee, cond-mat/0207655.

[30] J.P. Fagnard, P. Vanderbemden, D. Crate, V. Misson, M. Ausloos and R. Cloots, *Physica C* **372** (2002) 970.

[31] P. Vanderbemden, R. Cloots, M. Ausloos, R.A. Doyle, A.D. Bradley, W. Lo, D.A. Cardwell and A.M. Campbell, *IEEE Trans.Appl.Supercond.* **9** (1999) 2308.

[32] M.B. Maple, B.J. Taylor, N.A. Frederick, S. Li, V.F. Nesterenko, S.S. Indrakanti and M.P. Maley, *Physica C* **382** (2002) 132.

[33] G.K. Gupta, S. Sen, A. Singh, D.K. Aswal, J.V. Yakhmi, E.M. Choi, H.J. Kim, K.H.P. Kim, S. Choi, H.S. Lee, W.N. Kang and S.I. Lee, *Phys.Rev.B* **6610** (2002) 4525.

[34] E. Cappelluti, C. Grimaldi and L. Pietronero, cond-mat/0211481.

[35] S.X. Dou, X.L. Wang, J. Horvat, D. Milliken, A.H. Li, K. Konstantinov, E.W. Collings, M.D. Sumption and H.K. Liu, *Physica C* **361** (2001) 79.

[36] IAEA, "Compilation and evaluation of fission yield nuclear data", IAEA-TECDOC-1168 (2000).

[37] R.L. Heath, "Gamma-ray spectrum catalog", ANCR-1000-2.

[38] Y. Zhu, Z.X. Cai, R.C. Budhani, M. Suenaga and D.O. Welch, *Phys. Rev. B* **48** (1993) 6436.

[39] J. Provost, Ch. Simon, M. Hervieu, D. Groult, V. Hardy, F. Studer and M. Toulemonde, *Materials Research Society Bulletin* **20** (1995) 22.

[40] G. Szenes, *Phys. Rev. B* **51** (1995) 8026.

[41] F. Seitz and J.F. Koehler, in *Solid State Physics: Advances in Research and Applications,* edited by F. Seitz and D. Turnbull, Academic, New York, 1995, vol. 2, p. 305.

[42] M. Toulemonde, C. Dufour and E. Paumier, *Phys. Rev. B* **46** (1992) 14362.

[43] L. Civale, A.D. Marwick, M.W. McElfresh, T.K. Worthington, F.H. Holtzberg, J.R. Thompson and M.A. Kirk, *Phys. Rev. Lett.* **65** (1990) 1164.

[44] H. Safar, J.H. Cho, S. Fleshler, M.P. Maley, J.O. Willis, J.Y. Coutler, J.L. Ullman, P.W. Liowski, G.N. Riley, M.W. Rupich, J.R. Thompson and L. Krusin-Elbaum, *Appl. Phys. Lett.* **67** (1995) 130.

[45] J.F. Ziegler and J.P. Biersack, *The Stopping and Range of Ions in Solids,* Pergamon Press, New York, 1985.

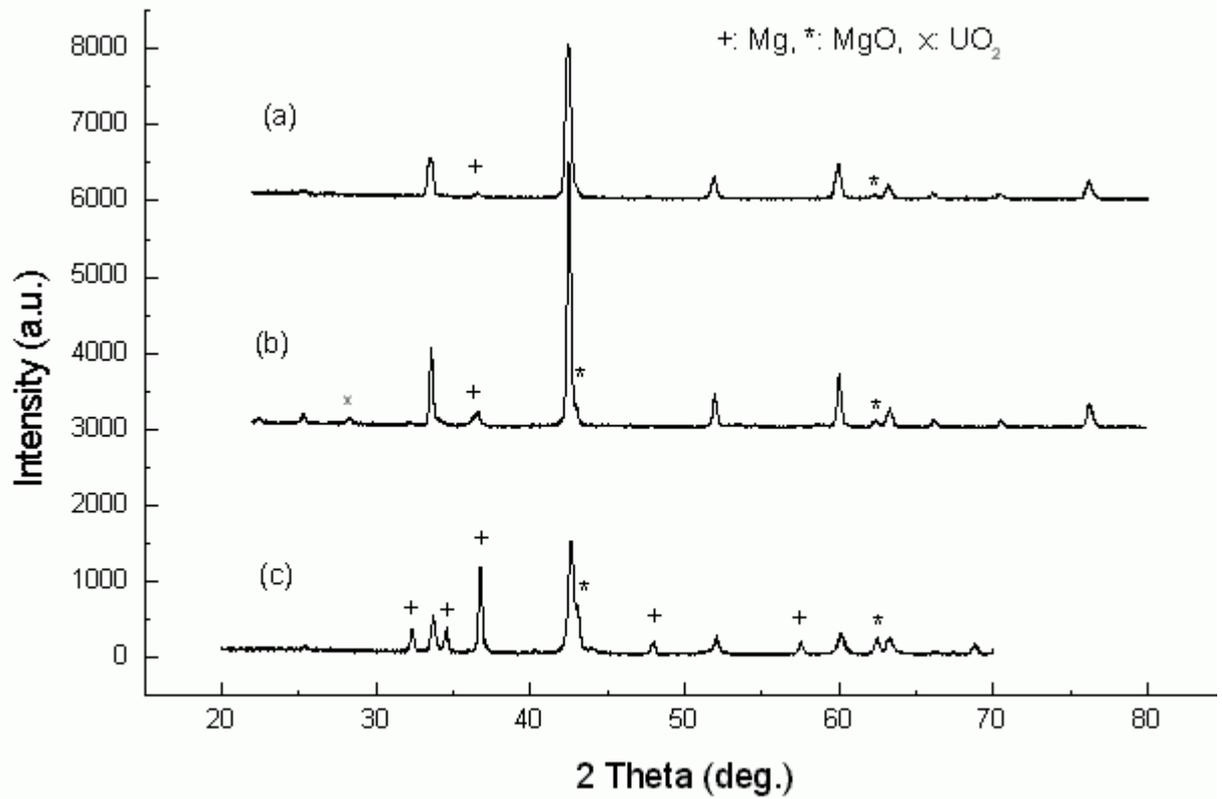

Fig. 1

Figure 1. XRD spectra of the MgB$_2$ used for the samples. (a) undoped, sintered at 900 C for 3h; (b) doped with 1 wt% U as UO$_2$, sintered at 900 C for 3 h; (c) undoped, sintered at 760 C for 0.5 h. The MgB$_2$ phase peaks are unmarked.

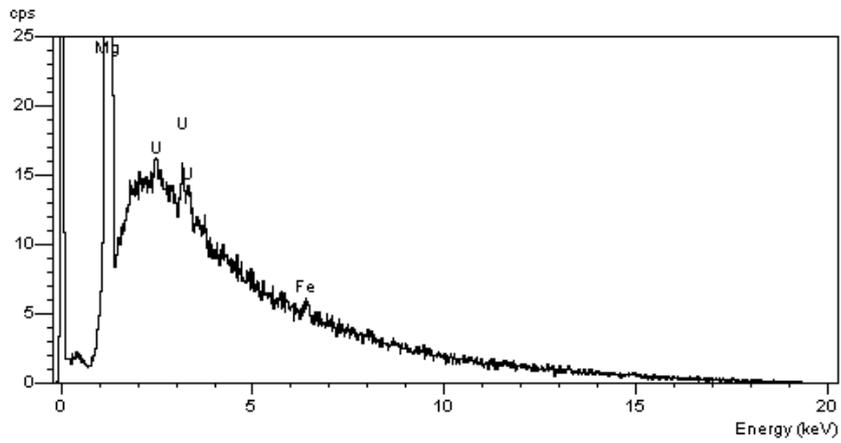

a)

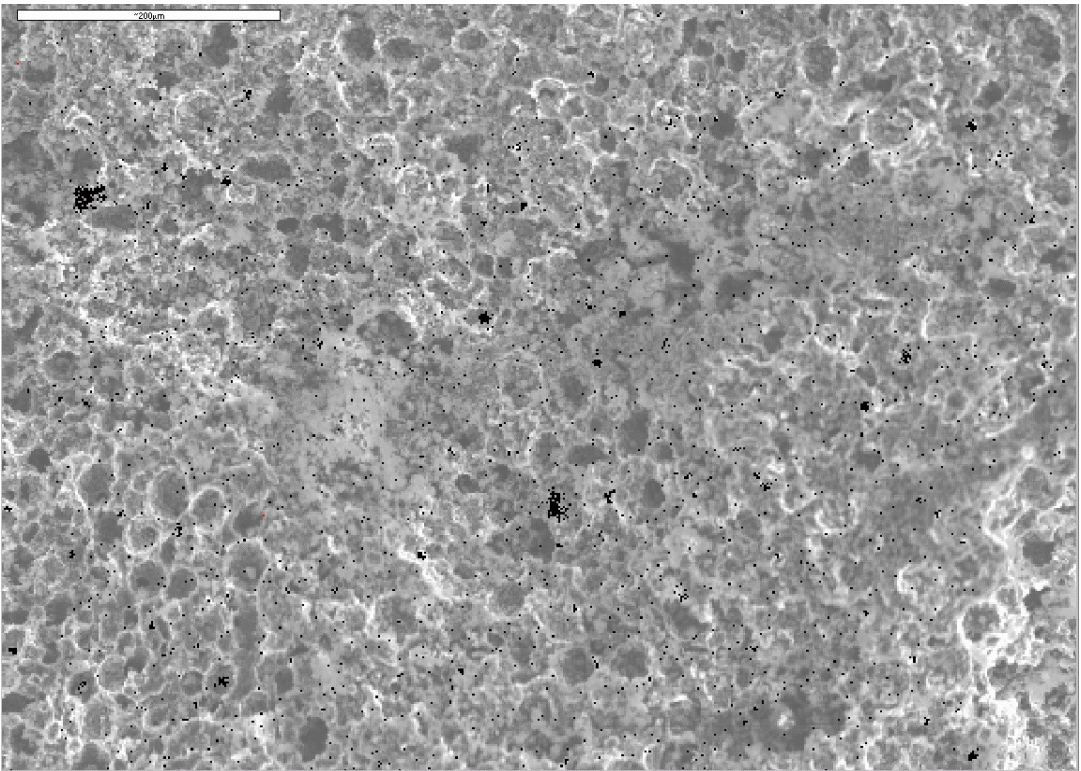

b)
Figure 2. (a) Averaged EDS spectrum of sample E doped with 1 wt % uranium; (b) EDS mapping of a cross-sectional area of sample E showing the distribution of the $UO_2$ particles in black.

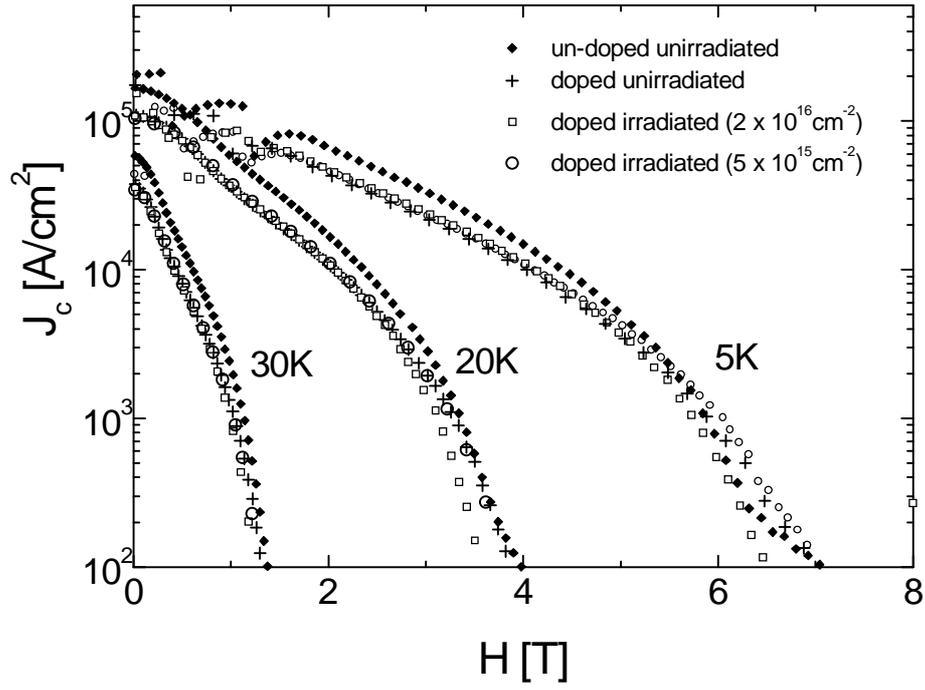

Fig 3 a)

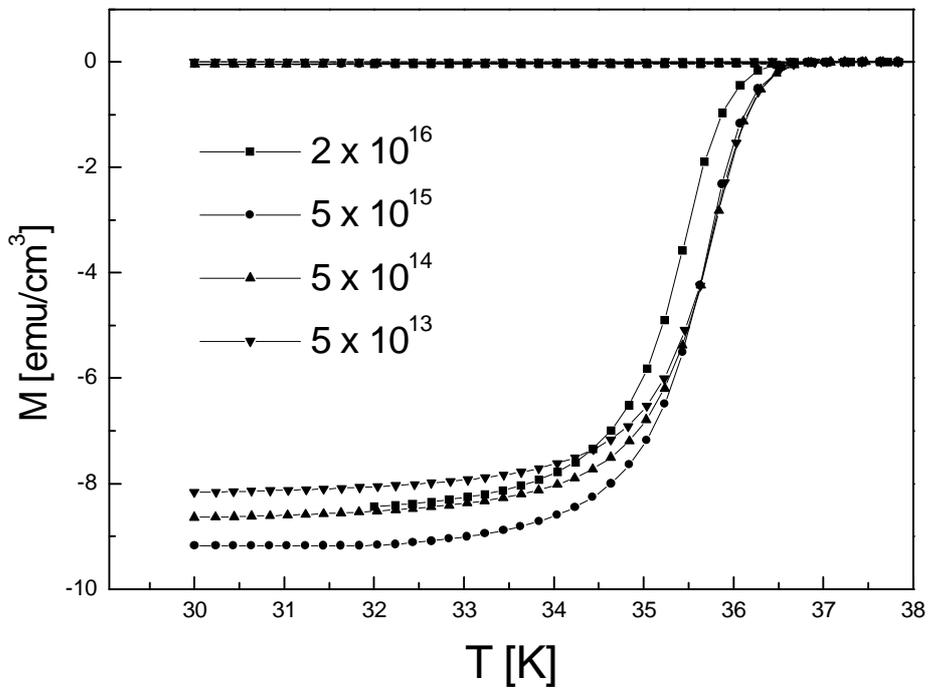

Fig 3b)

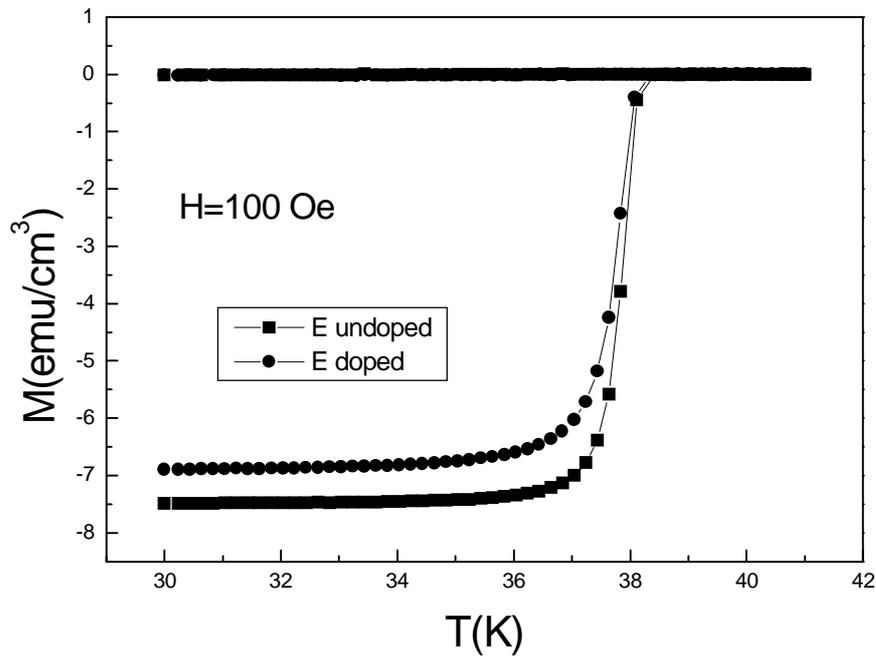

Fig. 3(c)
Figure 3. (a) Critical current derived from PPMS measurements as a function of magnetic field at 5, 20 and 30K for sample A (un-doped and doped with 1 wt% U). $J_c$ for the doped version of sample A is shown before and after irradiation at $2 \times 10^{16}$ cm$^{-2}$ with thermal neutrons. $J_c$ for the doped version of sample B (also 1 wt% U) is shown after irradiation with thermal neutrons at $5 \times 10^{15}/$cm$^2$. (b) The magnetisation under ZFC and FC conditions as a function of temperature at a field of 100 Oe as determined by MPMS measurements. All the samples shown were doped to 1 wt% U and irradiated at the fluences /cm$^2$ that are shown in the legend. (c) MPMS measurements under the same conditions for the un-irradiated high purity sample E, both un-doped and doped to 1 wt% U.

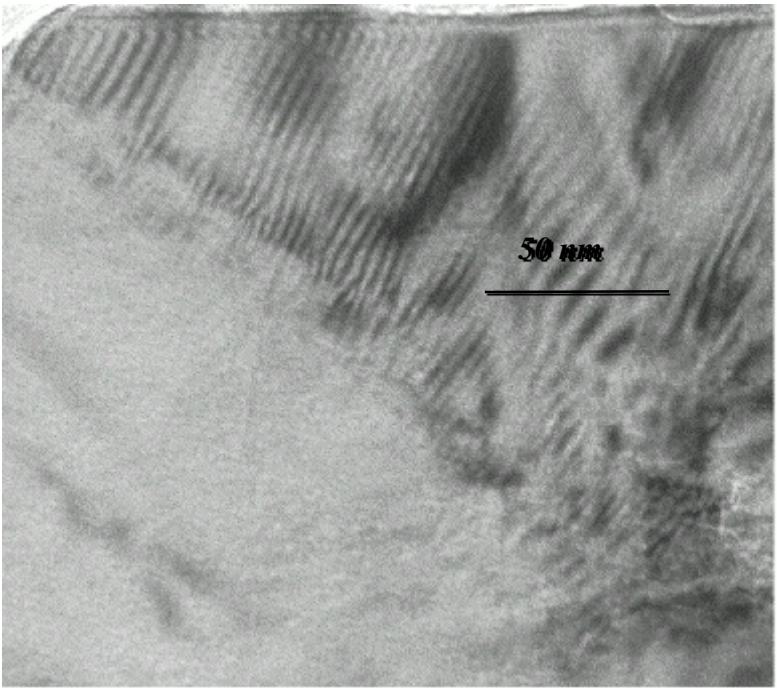
Fig. 4(a)

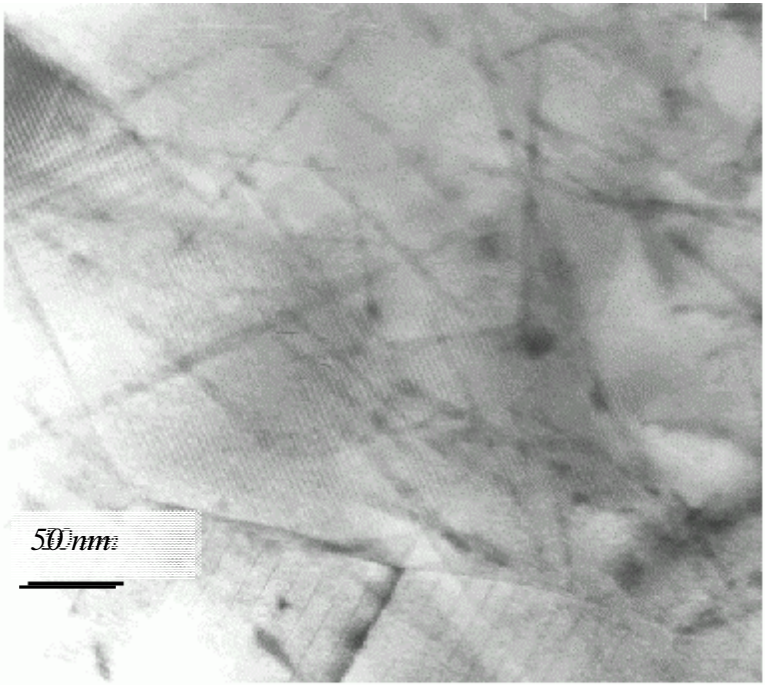
Fig. 4(b)

Figure 4. (a) Typical TEM image of an $MgB_2$ grain from sample B (doped to 1 wt% U and irradiated at $5 \times 10^{15}$ cm$^{-2}$ with thermal neutrons). This grain, like all the others examined, shows no evidence of columnar defects that might be due to fission tracks. (b) TEM image of the core of a U-doped irradiated Bi-2223/Ag tape showing fission tracks.

**Table 1:** Comparison of $^{235}$U fission yield estimated prior to irradiation and measured after irradiation using the gamma-ray spectrometric technique.

| Sample | $^{235}$U fission yield (Calculated) | $^{235}$U fission yield (Gamma spectrometric method) |
|---|---|---|
| C | $1.93 \times 10^{11}$ | $1.06 \times 10^{11}$ |
| D | $2.31 \times 10^{10}$ | $3.04 \times 10^{10}$ |